# More on random-lattice fermions

T.D. Kieu [a] [b], J.F. Markham [a] and C.B. Paranavitane [a]

[a] School of Physics, University of Melbourne, Parkville Vic 3052, Australia

[b] School of Natural Sciences, Institute for Advanced Study, Princeton NJ 08540, USA

The lattice fermion determinants, in a given background gauge field, are evaluated for two different kinds of random lattices and compared to those of naive and wilson fermions in the continuum limit. While the fermion doubling is confirmed on one kind of lattices, there is positive evidence that it may be absent for the other, at least for vector interactions in two dimensions. Combined with previous studies, arbitrary randomness by itself is shown to be not a sufficient condition to remove the fermion doublers.

## 1. Introduction

In a previous study [1] the fermion doubling problem was investigated on a special kind of random lattices. It was found on that lattice that the fermion doubling still existed if gauge invariance was always maintained for non-zero lattice spacing, even though random lattices are outside the catchment of the Nielsen-Ninomiya theorem. In the expectation that such negative results would be applicable to others kinds of random lattices, we now repeat the study for two more different kinds of random lattices.

The results to be presented here do not quite conform to our expectation: in one case there are doublers and none in the other in the continuum limit. The positive result is not a proof that certain kind of randomness will solve the doubling problem. To resolve the issue, detailed study in higher dimensions with non-abelian gauge groups will be necessary.

## 2. Random-lattice fermions

Two types of random lattices are considered. One is the random block lattice (RBL) [2] where the orthogonal structure of a hypercubic lattice is kept but with link lengths uniformly randomised from site to site and direction to direction. A typical two-dimensional RBL and its dual construc-tion is shown in figure 1. Each dual-lattice link is the bisector of a link of the original lattice, and each dual lattice cell confines one unique vertex of the original. A lattice of the other type considered and its dual is illustrated in figure 2. We call this type of lattices CFL after its advocators Christ, Friedberg and Lee [3, 4].

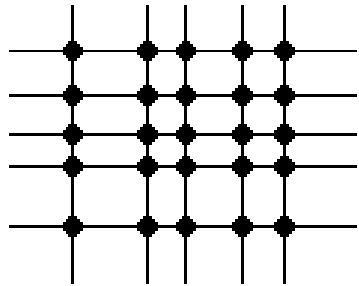

Figure 1. A fragment of random block lattice.

On both types of lattices the fermion action employed is

$$\mathcal{S} = \sum_\mu \sum_{ij} \bar\psi_i \frac{\sigma_{ij}}{l_{ij}} l^\mu_{ij} \gamma^\mu U_{ij} \psi_j + m \sum_i \omega_i \bar\psi_i \psi_i,$$

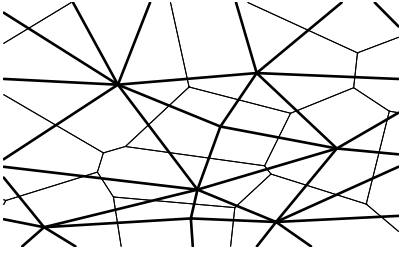

Figure 2. A fragment of CFL lattice and its dual lattice in thinner lines.

$$\equiv \sum_{ij} \bar{\psi}_i M_{ij}(U) \psi_j \,. \qquad (1)$$

The lattice sites are labeled by the indices $i, j$; the direction $\mu$ ($= 1, 2$); lattice link length $l_{ij}$ and the $\mu$-component of the link vector $l^\mu_{ij}$. The length of the dual-lattice link bisecting $\vec{l}_{ij}$ is $\sigma_{ij}$, while $\omega_i$ is the volume of the dual cell that contains the site $i$. $m$ denotes the fermion mass. The gauge link $U_{ij}$ is

$$U_{ij} = \exp\left\{ i \int_i^j A_\mu dx_\mu \right\} . \qquad (2)$$

In a square of length $L$ (in some arbitrary units) with opposite edges identified, we chose the abelian background gauge field to be

$$A_\mu = \delta_{\mu 1} \frac{EL}{2\pi} \cos\left(\frac{2\pi x_2}{L}\right). \qquad (3)$$

### 3. Results and Remarks

For a measure of the number of fermions species we evaluate the quantity introduced in [1]

$$\Gamma \equiv \ln\left\{ \det[M(0) M^{-1}(U)] \right\}. \qquad (4)$$

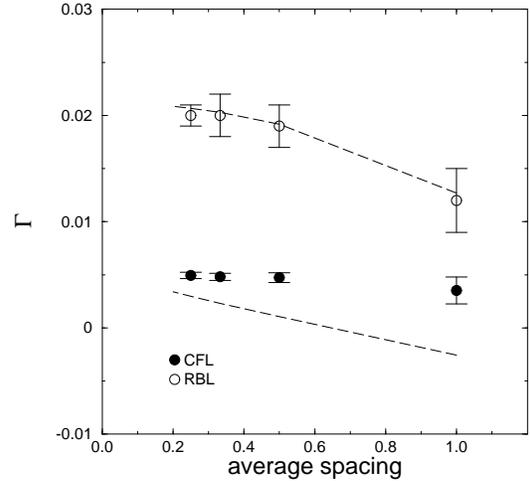

Figure 3. The continuum limit of different lattice fermions in a given background gauge field.

This contains the various fermion one-loops in the bacground gauge field configuration; and the number of fermion species in the loops can be extracted from an overall integral factor. We compare the random-lattice fermions to naive and wilson fermions in the continuum limit with fixed physical quantities $m_{\text{phys}} = 0.4$, $E_{\text{phys}} = 0.05$ and $L = 8.0$ given in some arbitrary units. The results are presented in figure 3. The data points are obtained after averaging over ensembles of 30 random lattices for each average lattice spacing of the four values 1.0, 0.5, 0.333 and 0.25. The upper and lower dashed lines represent the approach to the continuum limit for naive and wilson fermions, respectively.

It is evident from our numerical results that the fermion doubling problem is persistent on the random block lattices, which follow the behaviour of naive fermions on regular lattice, which has fourfold doubling. Thus despite the good continuum behaviours of suitably defined lattice operators studied elsewhere, this kind of lattices is not appropriate for the removal of doublers.

In contrast, the CFL lattices exhibit no doubling behaviour. They approach the wilson fermion limit. If this positive result persists in higher dimensions, which we expect it would, we may have a superior lattice regularisation over the wilson fermions. The continuum space-time invariance is better approximated by ensemble of random lattices and the scaling window could be broader than on regular lattices, as illustrated in the graph above where the approach to continuum limit is very flat for the CFL lattices. (However, there are claims otherwise, see [5], perhaps because a single random lattice is investigated there one at a time instead of an ensemble of lattices.)

All these gains aside, the most direct approach to implement (anomaly-free) chiral gauge theory on the CFL lattices is apparently not gauge invariant. In order to have the doublers decoupled, an explicit fermion mass term needs to be introduced and the massless limit should only be taken after the vanishing lattice spacing limit [4]. The behaviour of massless fermions on CFL lattices is not yet fully understood. With vector interactions, the determinant test, as above, indicated a trend of no doubling similar to the wilson fermions; but the anomalous current divergence could not be reproduced [6]. Some point-split formulation of chiral fermion currents in the action chiral gauge theory may be necessary; non-locality could thus creep in and even so chiral gauge invariance cannot be guarantee [7]. However, more sophisticated approach to chiral gauge theory on random lattices whereby the Higgs mechanism is invoked remains a feasibility. We hope to come back to this last approach elsewhere.

We also want to understand the mechanism, perhaps related to the localisation in disordered medium, that triggers the removal of doubling on a kind of random lattices but not the others. Arbitrary randomness by itself is clearly not a sufficient condition for no fermion doubling even though it violates the assumptions of the lattice Nielsen-Ninomiya theorem. This conclusion is also supported by a study of doped lattice [8].

**Acknowledgements**

TDK wishes to thank Ting-Wai Chiu, Hebert Neuberger and R. Narayanan for discussions. He also acknowledges the hospitality of Steve Adler and the Institute for Advanced Study during a working visit, and the support of the Australian Research Council.